# Effect of Anisotropic Peierls Barrier on the Evolution of Discrete Dislocation Networks in Ni

John D. Shimanek[1], Darshan Bamney[2], Laurent Capolungo[2], Zi-Kui Liu[1], Allison M. Beese[1,3]

[1] Department of Materials Science and Engineering, The Pennsylvania State University, University Park, PA 16801, USA

[2] Materials Science & Technology Division, Los Alamos National Laboratory, Los Alamos, NM 87544, USA

[3] Department of Mechanical Engineering, The Pennsylvania State University, University Park, PA 16801, USA

Abstract

Over low and intermediate strain rates, plasticity in face centered cubic (FCC) metals is governed by the glide of dislocations, which manifest as complex networks that evolve with strain. Considering the elastic anisotropy of FCC metals, the characteristics of dislocation motion are also anisotropic (i.e., dislocation character angle-dependent), which is expected to notably influence the overall evolution of the dislocation network, and consequently, the plastic response of these materials. The aggregate influence of the anisotropy in the Peierls stress on the mechanical response of single crystal Ni was investigated in the present work using discrete dislocation dynamics simulations. Twenty initial dislocation networks, differing in their configuration and dislocation density, were deformed under uniaxial tension up to at least 0.9% strain, and the analysis of character-dependent dynamics showed a suppression of plasticity only for segments of nearly screw character. While the increased screw component of the Peierls stress raised the initial strain hardening rate, it also resulted in longer dislocation segments overall, contrary to the reasoning that longer pinned segments exhibit a lower resistance to motion and might give a weaker response. A non-linear superposition principle is demonstrated to predict the hardening reasonably well, considering the cumulative effects of forest and Peierls stress-related strengthening. Further analysis of the network topology revealed a tendency to maintain connectivity over the course of deformation for those networks simulated using an unequal Peierls stress. The general increases in hardening rate and network connectivity contrast with the localized reduction of dislocation motion, which occurred mainly for segments of nearly screw-type character.

<u>Introduction</u>

At low and intermediate deformation rates, plasticity in face centered cubic (FCC) metals is accomplished by the glide of dislocations, especially for materials with at least moderate stacking fault energies [1]. The material's bulk plastic behavior, including yield and strain hardening rate, are related to the evolution of complex networks of dislocations that develop to accommodate plastic deformation [2]. With the evolution of dislocation networks depending on the motion of individual dislocation segments, much work has been focused on the details of the dislocation core properties and how they affect resistance to deformation [3]. In particular, the core structure of dislocations and the associated behaviors (i.e., dynamics, interactions) are strongly dependent on the dislocation character, which is defined by the relative orientations of the dislocation line and its Burgers vector. Unraveling the overall effect of character dependence on the friction stress of dislocations is the focus of the present work.

For FCC metals, the classical view of dislocation dynamics has typically been framed in terms of their negligible Peierls barrier and the resulting importance of dislocation interactions to the overall strain hardening rate [4]. However, the shear stress required to move a dislocation, interchangeably called the Peierls stress or lattice friction stress, has been shown experimentally to be important, at least in pure Ni at higher pressures [5]. In the same material system at zero hydrostatic pressure, recent estimates using a modified Peierls-Nabarro equation and input from calculations based on density functional theory gave a highly anisotropic friction stress, ranging from 9.4 MPa for pure edge to 300 MPa for pure screw [6]. Separately, nudged elastic band calculations of the Peierls energy barrier show similarly large spreads across dislocation characters for pure Ni, with screw components having energies higher by a factor of around 40 [7], noting that this ratio is of energies, not stresses.

The character dependence of dislocation motion appears also in the behavior of other FCC elements. Anisotropic elasticity considerations combined with energy values from the generalized stacking fault surfaces of Cu, Ni, and equiatomic FeNiCoCrMn showed a strong and sometimes irregular character dependence on the calculated friction stresses [8]. An atomistic study of the effect of normal stress on the Peierls barrier to dislocation glide in Al showed a relatively large value for the Peierls stress of a screw component [9]. Additionally, the effect of anisotropy was more noticeable nearer the screw components than the edge [9]. The mobilities of dislocations in FCC materials have also been shown to depend on character, as reported from molecular dynamics simulations of both pure Al and pure Ni [10,11]. For micromechanical simulations on the level of grains, a character based partitioning of dislocation populations has been used to account for the ability of edge dislocations to climb and for screw dislocations to cross slip within a dislocation density based crystal plasticity hardening law [12], although the Peierls barrier was treated as an isotropic value.

Exploring the mesoscale interactions of two-dimensional crystalline defects and their influence on bulk material strength has long been a goal of discrete dislocation dynamics (DDD) [13]. The computational expense of direct numerical simulations of highly interactive dislocations systems has typically limited DDD results to relatively low cumulative strains, but achievable strain levels have been steadily increasing with the power of computational resources and with the reformulation of the mechanical problem into the cast of efficient spectral methods [14,15]. Simulated strains of one or two percent allow for the investigation into representative periods of plastic activity, enabling connections to be made between the atomistic origins of dislocation motion and their eventual effect on the bulk response during plastic deformation. With these recent advancements, DDD is well positioned to explore how dislocation properties originating from atomistic behavior scale up to affect network evolution and overall bulk material response.

Previous DDD studies on FCC materials related the total density of dislocations to the flow stress [16], which took the form of a Taylor-like square root dependence [17]. Such simulations are in apparent agreement with a wealth of experimental data [18], yet the focus of modeling efforts has mainly been on the ability of junction properties to connect the length scales of DDD and homogenized crystal plasticity [19,20]. However, a direct connection from DDD results to crystal plasticity remains incomplete due to the complexity of the relevant physical mechanisms.

In light of the potential for anisotropic Peierls stresses to exist within pure Ni, as noted in Refs. [6,13–15], the present study focuses on the evolution of dislocation networks through the lens of dislocation character. Specifically, the objectives of this work are to investigate the influence of character-dependent Peierls stress on the mechanical response of synthetic dislocation microstructures, with analysis relating the bulk response to the underlying length and connectivity of the dislocation segments within the networks. To this end, several statistical volume elements of dislocation networks were evolved under uniaxial tension under cases of both equal and unequal Peierls stress for edge and screw components. The lengths and swept areas of all dislocation segments were tracked throughout deformation to provide character-based measures of mean free path lengths. The lengths of contiguous segments were also tracked due to their importance to the overall strength response in many theoretical frameworks of crystal plasticity. The analysis suggests that, despite the localized reduction in plasticity for segments with the highest Peierls stresses, the global effect of the applied friction stress anisotropy on dislocation network evolution was significant. Furthermore, the strengthening effect was captured well with a superposition law of dislocation and Peierls stress contributions.

Methods

*Discrete Dislocation Dynamics*

The DDD method used here is described in Refs. [21,22]. Briefly, the underlying micromechanical framework is a spectral approach coupled with the field dislocation mechanics method to efficiently calculate the long-range mechanical fields associated with ensembles of dislocations embedded in an inhomogeneous continuum [23]. Further, near-core corrections are included, which allows for accurate descriptions of the elastic fields of dislocations in the short range in anisotropic media. The following notational scheme is adopted throughout this section: scalar quantities are represented by italicized characters, vectors are denoted by bold symbols, and second-order tensors are denoted by bold and italicized symbols.

In nodal DDD, an overdamped equation of motion is used to describe the relationship between the force acting on the dislocation nodes and their corresponding velocities. Forces at nodes arise due to the stresses experienced along the segments between nodes, which are represented in a linear form. Therefore, to find the motion of each node, the contribution from each segment is found as

$$\mathbf{f}_{ij} = l_{ij} \int_0^1 (1-s) \big[ \big( \boldsymbol{\sigma}(\mathbf{x}_{ij}) \cdot \mathbf{b}_{ij} \big) \times \mathbf{t}_{ij} \big] \, ds \qquad (1)$$

Here, the interpolation between nodal positions is linear along its length, represented as the integration variable $s$, and the length of the segment is $l_{ij} = \|\mathbf{x}_j - \mathbf{x}_i\|$, where $\mathbf{x}$ denotes spatial position. The Burgers vector for segment $ij$ is $\mathbf{b}_{ij}$ and the unit tangent vector is $\mathbf{t}_{ij}$.

The stress $\boldsymbol{\sigma}$ is a function of position and is the result of both internal short range elastic interactions and macroscopically imposed average loads.

The present interest lies in the imposition of character-dependent Peierls stresses, which represent the minimum stress required to move a dislocation. The dependence of Peierls stress on dislocation character was calculated during the simulation based on the following interpolation scheme, similar to that found for the character dependence of core energy for Al based on atomistic simulations [24]:

$$\tau_P(\phi) = \tau_P^e \sin^2 \phi + \tau_P^s \cos^2 \phi \qquad (2)$$

where $\phi$ is the character angle, $\tau_P$ is the Peierls stress, and the superscripts refer to edge or screw values prescribed before the simulation begins. Translating the stress to a segment force through Eqn. 1 gives

$$f_{ij}^P = \tau_P b_{ij} l_{ij} \qquad (3)$$

where $b_{ij}$ is the magnitude of the segment Burgers vector and $l_{ij}$ its length. Therefore, the total force on a node must be compared to the Peierls stress projected onto the slip plane. The effective nodal force considered for mobility purposes is then

$$\mathbf{f}'_{ij} = \begin{cases} \mathbf{f}_{ij} - |\mathbf{f}_{ij}^P| \cdot \left( \dfrac{\mathbf{f}_{ij}}{\|\mathbf{f}_{ij}\|} \right) & \text{if } \|\mathbf{f}_{ij} > \mathbf{f}_{ij}^P\| \\ 0 & \text{if } \|\mathbf{f}_{ij} \leq \mathbf{f}_{ij}^P\| \end{cases} \qquad (4)$$

The total force on each node is then the sum over connected segments

$$\mathbf{F}'_i = \sum_j \mathbf{f}'_{ij} \qquad (5)$$

where $\mathbf{F}'_i$ denotes the effective nodal force after the Peierls stress has been subtracted. Finally, the motion of each node, which determines the evolution of plastic strain, is given by a linear mobility law:

$$\mathbf{F}'_i = \sum_j \boldsymbol{B}_{ij} \mathbf{v}_j \qquad (6)$$

where $\mathbf{v}_i$ is the velocity of node $i$, $\boldsymbol{F}'_i$ is the total force acting on it, including short range interactions and the macroscopic stress state, and $B$ is the damping coefficient. The damping coefficient depends on dislocation character and may be obtained from atomistic simulations probing the relationship between applied shear stress and the velocity of an isolated dislocation [11].

With the addition of a character dependence to the Peierls stress, anisotropy is considered throughout the calculation of nodal velocities: short range forces accounting for anisotropic elasticity are summed for each node and any stress above the friction stress results in a velocity proportional to a character-based mobility constant.

*Simulation Setup*

The simulation cells were cubic with 1000 lattice parameters (352nm) on a side, each discretized with 64 Fourier grid points. The simulations were oriented so that the loading orientation was one degree away from a [001] crystallographic orientation along a double slip boundary. Loading directly along [001] results in the ability of segments traversing the periodic boundary conditions to directly interact with themselves, which usually resulted in spurious annihilation and a loss of dislocation density [16]. The tilted crystallographic orientation resulted in image segments that were able to interact only through their elastic fields, as discussed in the supplemental information. Based on previous first-principles calculations [6], the elastic constants and Peierls stress for pure Ni used in the DDD simulations are given in Table 1 along with linear damping coefficients at room temperature for the dislocation mobilities [11]. The Peierls stress calculations from Ref. [6] used first-principles calculations based on density functional theory to obtain both the ideal shear strength and the full elastic properties of pure Ni; those quantities were then translated through a Peierls-Nabarro framework to estimates of the minimum stress required to move a dislocation [25]. The calculations accounted for differences in dislocation character through analytical forms for edge and screw dislocations derived in Ref. [26], resulting in a larger estimate of the screw component Peierls stress.

Table 1. Material parameters used in the DDD simulations of pure Ni. All values were taken from [6], except for the damping coefficients, which were calculated for 300 kelvin based on data in [11].

| Parameter | Value | Units |
|:---:|:---|:---|
| $C_{11}$ | 265 | GPa |
| $C_{12}$ | 161 | GPa |
| $C_{44}$ | 127 | GPa |
| $a_0$ | 3.52 | Angstroms |
| $\tau_P^e$ | 9.4 | MPa |
| $\tau_P^s$ | 282 | MPa |
| $B^e$ | 1.5 | MPa·s |
| $B^s$ | 1.92 | MPa·s |

The approach to initialize dislocation networks followed that outlined in Refs. [27,28]. Briefly, each simulation cell was first populated with an equal number of circular dislocation loops on all slip systems. A substantial and complex stress state was then applied, with components on the order of 0.5 - 2 GPa for both normal and shear components. Following this, zero stress boundary conditions were imposed, and the dislocation ensemble was allowed to relax and reconfigure under the influence of internal elastic fields only. The change in dislocation density per dynamics time step was below 0.001% at the end of relaxation, and the final initialized states were examined for anomalies in Paraview [29]. In this way, 20 dislocation configurations were initialized, each using equal Peierls stress for edge and screw components. Copies were then made of each relaxed configuration, the Peierls stress was increased for screw-type segments, and those copies were relaxed for an additional amount of time. An example of one of the relaxed networks, which had an initial dislocation density of $1.0 \times 10^{15}$ m$^{-2}$, is shown in Figure 1. The distribution of initial dislocation densities, which range from $5.1 \times 10^{14}$ to $1.3 \times 10^{15}$ m$^{-2}$, is given in Figure 2.

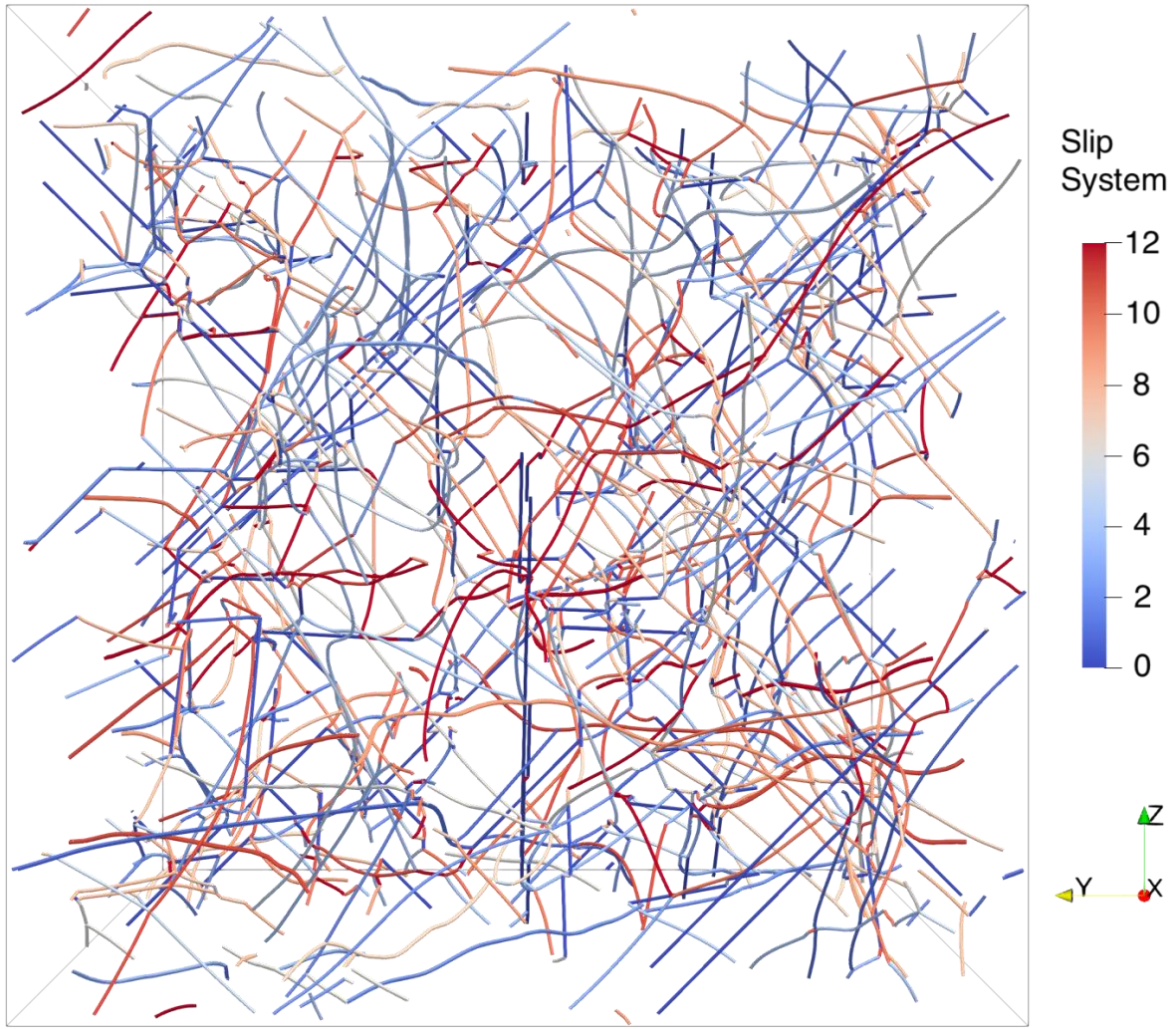

Figure 1. A snapshot of one of the relaxed dislocation networks, with segments colored by slip system. A slip system of 0 indicates a junction while all other systems are the 12 FCC systems in the family $\langle 001 \rangle \{11\bar{1}\}$.

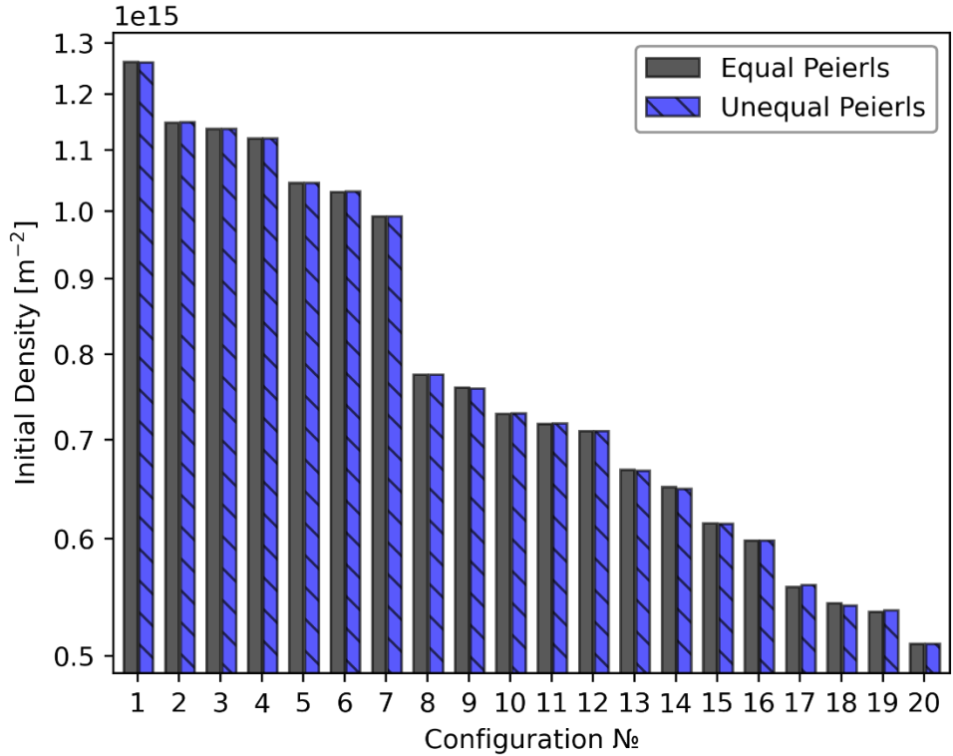

Figure 2. Overall distribution of dislocation densities among initialized dislocation networks, ordered from highest to lowest. Cases of equal and unequal Peierls stress originating from the same relaxed configuration are group together. Note that the dislocation densities are plotted on a log scale.

After relaxation, macroscopic loads were imposed to the simulation domains via a quasistatic loading scheme, which is discussed further in the supplemental information. The stress components for all components other than the axial component, which was strain controlled, were set to zero to simulate uniaxial tension.

*Output Data*

Several types of dislocation data were extracted for the examination of character angle-based statistics, including, e.g., the length of total dislocation segments having a certain character. Additionally, the plastic area swept by each dislocation segment quantifies the contribution from each segment to the overall plasticity within a network. Together, length and area also provide an estimate of the mean free path traveled as a function of character angle, which will be discussed later.

It is important to note that the length metrics produced in the above give no indication of the number of dislocation segments. Particularly, the total dislocation line length normalized by the simulation cell volume provides only an average measure of the dislocation density, and is insufficient to answer the question of how this total length is partitioned into segments of different attributes. Comparisons to theory would be aided by counting and measuring the length of contiguous segments and so to know, for instance, whether a system's total dislocation density was composed of many small, entangled dislocation segments, or only several large segments with very little contact with each other. To count segments, the dislocation network was navigated according to its connections, with information about one segment being accumulated until reaching the next connection point. This amounts to a depth-first search of the network with each contiguous segment represented as a node of the constructed graph. Then, the connections among nodes correspond to edges within the graph representation adopted here. The NetworkX Python package [30] is used for the graph analysis described in the next section. An example of this representation for the process of junction formation is illustrated in Figure 3.

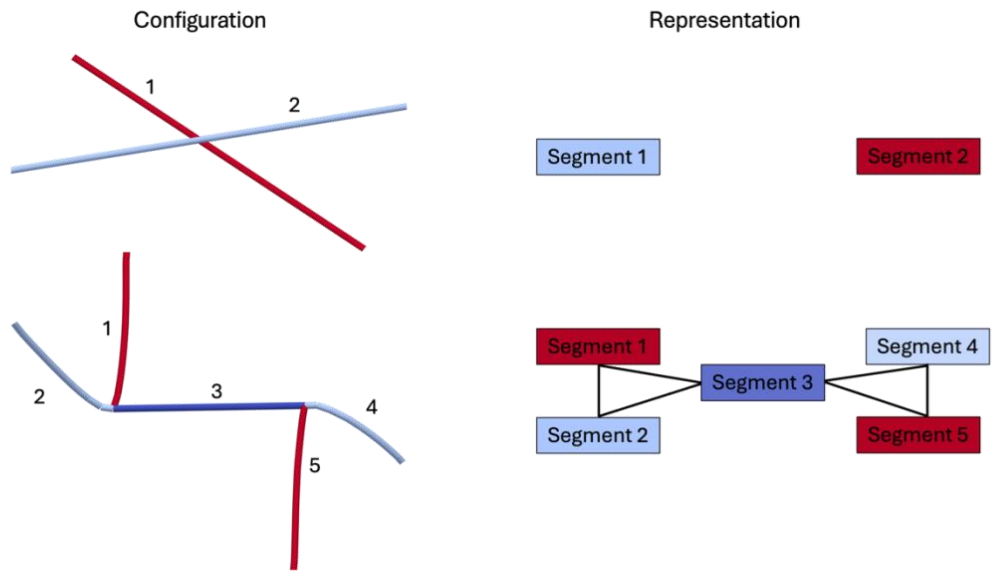

Figure 3. Two dislocation configurations and their representation used for statistics on contiguous segments. Contiguous segments on the left are recorded as nodes (boxes) on the right, with edges indicating their connection. The top row shows a junction just about to form, resulting in a representation of two components. Once formed, the junction segment brings all the segments into one connected component.

Results

*Stress-Strain Response*

Given the initialization strategy employed here, the simulations can be grouped into pairs that begin deformation with comparable configurations (i.e., comparable dislocation densities). An example of the stress-strain response from one such pair is shown in Figure 4; this same pair, labelled as 6 in Figure 2, will be used throughout as a representative example for discussion. For this configuration, the increased Peierls stress (termed "unequal case") resulted in a higher hardening response overall, with the response of the original material parameters (termed "equal case") showing earlier signs of plastic activity.

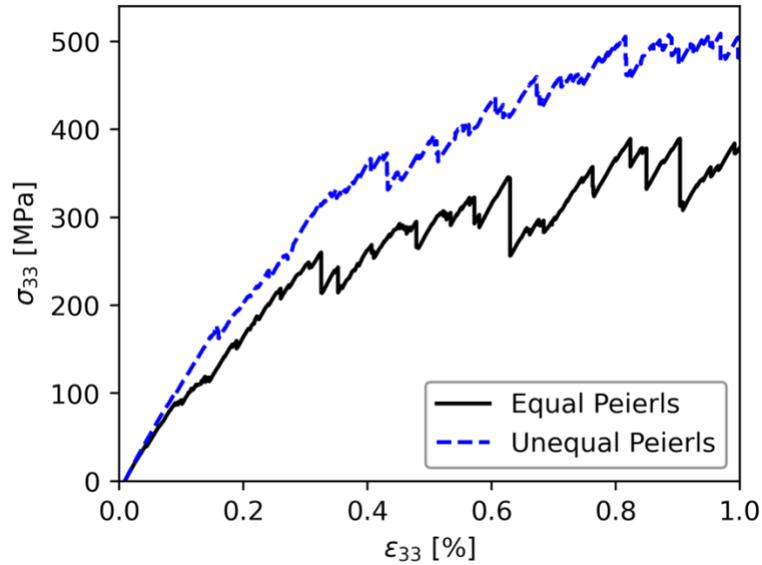

Figure 4. Stress-strain behavior of two comparable initial dislocation configurations (labelled as configuration 6 in Figure 2) with material properties differing only in the higher Peierls stress for screw-type dislocations.

Each of these simulations is best regarded as a statistical rather than representative sampling of the possible responses of the material depending on the initial dislocation configuration. Therefore, the average difference in the stress-strain curves was calculated. A fixed strain range, from 0.4% to 0.9%, was chosen to fairly compare the initial hardening rate among configurations. This section was linearly fit after being binned into points of equal spacing along strain. The initial elastic region was also fit, and its extrapolated intersection with the initial hardening slope gave a representation of approximate yield location, although significant plasticity on the scale of the individual dislocations typically occurred before such a point. The mean initial hardening rates for both equal and unequal cases, averaged across all configurations, are given in Table 2 and show a significant increase for the case of increased screw component Peierls stress.

Table 2. Mean and standard deviation of initial hardening rates for all dislocation configurations for each of the Peierls stress cases.

| Case | Mean hardening rate | Standard deviation |
|---|---|---|
| $\tau_P^s = \tau_P^e$ | 86 MPa | 80 MPa |
| $\tau_P^s = 30\,\tau_P^e$ | 190 MPa | 72 MPa |

*Character-Based Statistics*

The length and swept area of dislocations were tracked over the course of deformation. Of primary interest is the distribution of length and area across dislocation characters, which is shown in Figure 5 for the example configuration (labeled as 6 in Figure 2) for the unequal case, averaged over strains ranging from 0.4% to 0.9%. There, several peaks are evident in the distributions: near both pure edge and pure screw segments and around 60˚. The length distribution shows a larger peak around screw segments than the area distribution, indicating a higher total length of near-screw segments whose plastic activity was not as significant. This is in agreement with the applied increase in Peierls stress for screw segments, decreasing their overall mobility. Indeed, the area distribution as a whole tends to favor edge-type segments. The over-representation of near 60˚ dislocation segments was not due to any crystallographic reasons but rather to the effect of periodic boundary conditions with the specific magnitude of tilt away from the [001] loading axis. This is discussed more in the supplemental information.

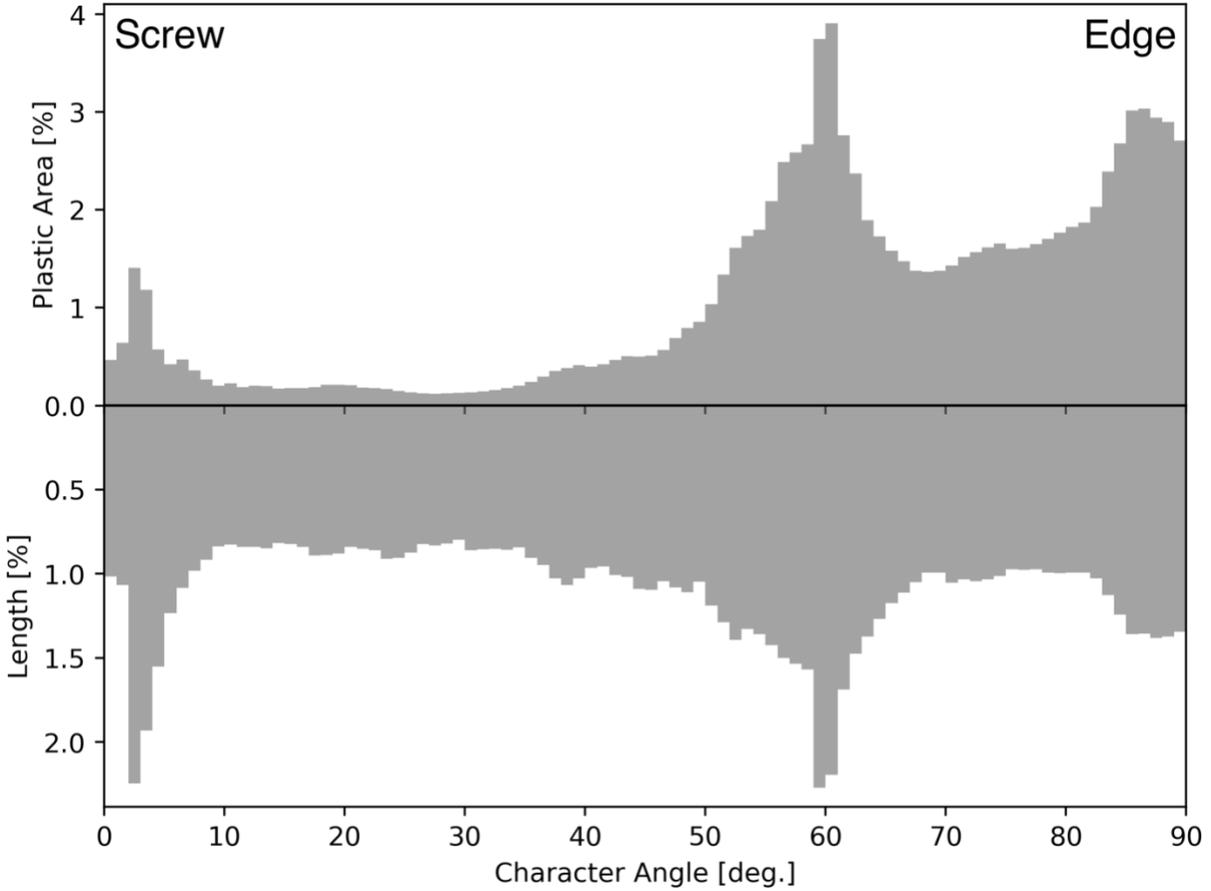

Figure 5. Relative distributions of plastic area (top) and length (bottom) over dislocation character, averaged over an axial strain of 0.5%. The initial configuration was labeled as 6 in Figure 2 with a high Peierls stress for screw components.

As a measure of plasticity, the area distribution cannot be easily interpreted due to its dependence on the length distribution. To lessen the confounding role of configuration, the ratio of swept area to length of the segment can instead be evaluated as an estimate of the mean free path, i.e.,

$$\Lambda(\theta) = \sum_i \frac{A_i(\theta)}{L_i(\theta)} \tag{7}$$

where $\Lambda$ is the mean free path, $A$ is the swept area, $L$ is the dislocation length, and they are all functions of the dislocation character, $\theta$. The values are summed over a representative period of plasticity available for all simulations, individual points of which are indicated by the subscript $i$, ranging from 0.4% to 0.9%. The estimate of mean free path is shown in Figure 6.

Given the importance of configuration, another quantity of interest is the difference of this estimated mean free path from what might be expected in any given configuration. If dislocations of all characters are presumed to accomplish the same amount of plasticity, then the mean free path distribution should follow the length distribution. The difference from that isotropic estimate and the observed distribution then represents the excess mean free path, $\Lambda - \langle L \rangle$, for which only the dynamics (and not the configuration) are responsible. Here, the mean length of all $N$ segments in the network is still a function of character:

$$\langle L \rangle = \langle L(\theta) \rangle = \frac{1}{N} \sum_i L_i(\theta) \tag{8}$$

Since the configuration is accounted for by the length distributions and all simulations were run over the same range, the excess mean free path is comparable between runs and its average is shown in Figure 7 for cases of both equal and unequal Peierls stress simulations.

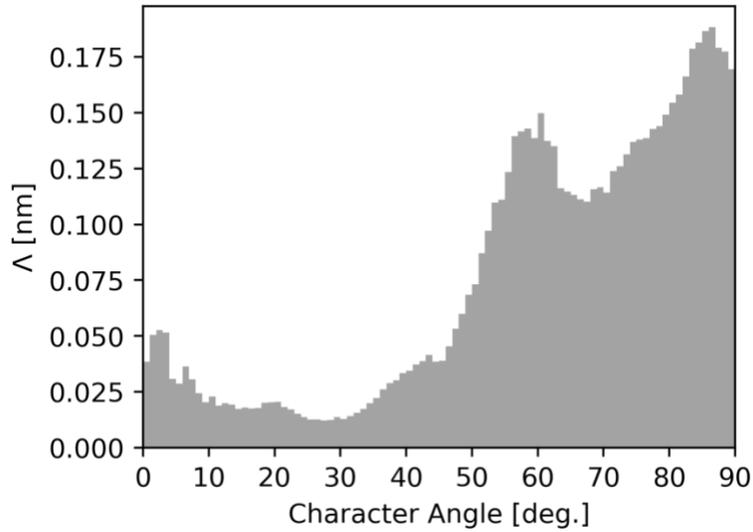

Figure 6. The ratio of area swept over dislocation length used as an estimate of the mean free path, $\Lambda$, given as a function of character angle for the same configuration 6 (defined in Figure 2) with high screw Peierls stress as shown in Figure 5.

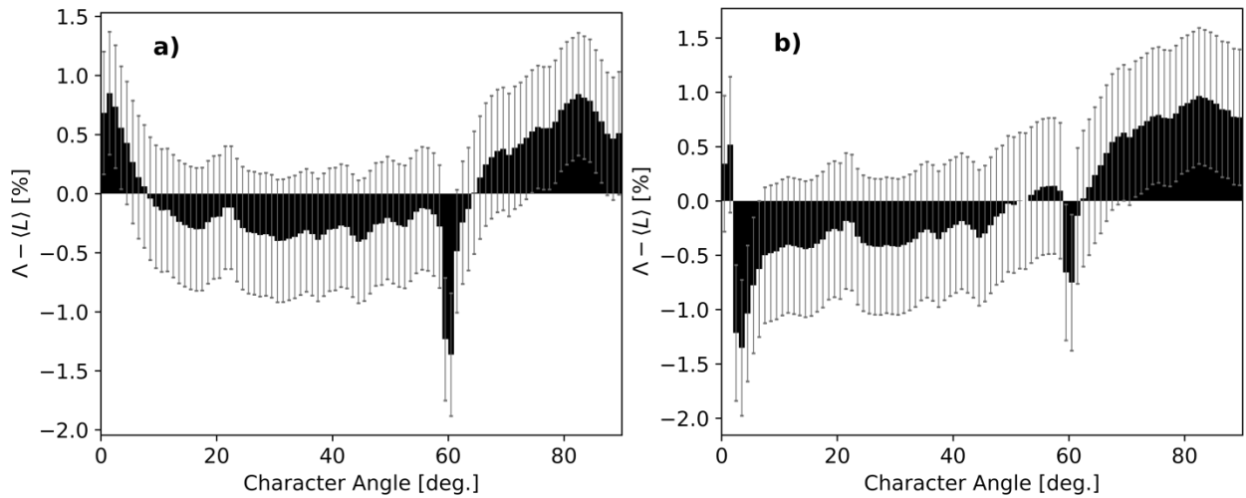

Figure 7. Excess mean free path distribution for the case of a) equal Peierls stresses and b) high screw Peierls stress. Mean free path, $\Lambda$, above what is expected based on prevalence in the network, $\langle L \rangle$, is taken as a comparable measure of dynamics. Black bars represent the mean of this value across simulations, with error bars indicating standard deviation.

Despite the large increase in the imposed Peierls stress for screw segments, the decreased mean free path is rather localized to within around 10˚, indicating a relatively lower activity in the dynamics for those segment types. Beyond that region, the distributions, which are averages over all simulation runs, show remarkable similarity. Both show another depression of the mean free path around 60˚ for the same reasons regarding image segment dipole interactions previously mentioned. The existence of these dipoles is not expected to influence the present conclusions since unequal Peierls stress case inherited their initial configuration from the equal case, meaning the dipole population would be identical in both cases.

*Length Distributions*

Besides metrics tied to the dynamics of dislocations in the networks, further information can be gleaned about the structure of each configuration from static metrics describing the network. One such quantity of particular importance to crystal plasticity theory is the distribution of segment lengths. Often, this is related for simplicity to the average dislocation density as $L_\rho = 1/\sqrt{\rho}$, although some have noted the importance of spatial heterogeneity [31]. Here the subscript indicates that $L_\rho$ is a homogenized estimate of line length based only on the dislocation density. Figure 8 compares the effect of Peierls stress on the evolution of average length for the example configuration, showing a steep increase for the case of equal Peierls stress relative to the unequal case. Note that for measures of length describing the size of individual segments, the symbol $L$ is used; for other cases of length whose exact definition will be discussed, the symbol $l$ is used.

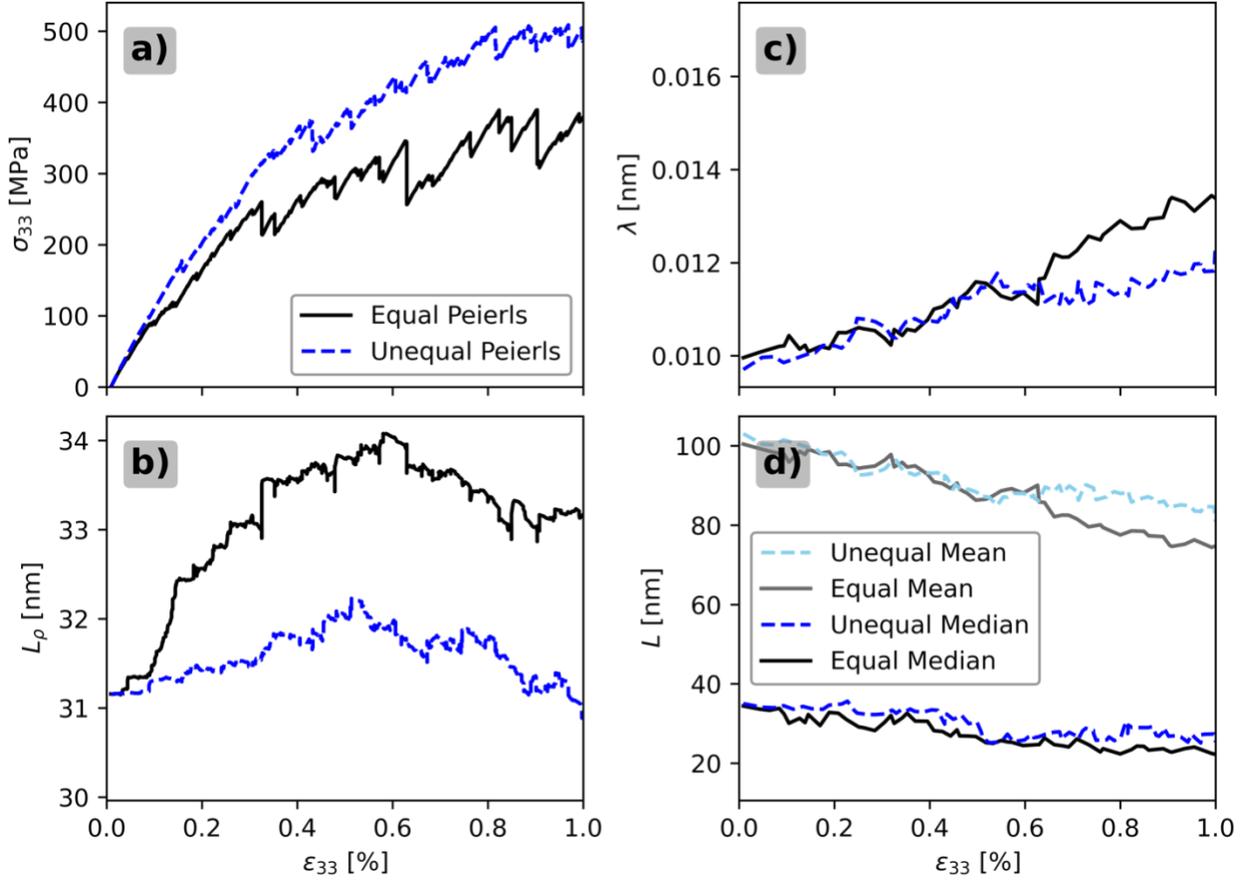

Figure 8. A comparison of the equal and unequal Peierls stress cases for configuration 6 (defined in Figure 2) showing a) the stress-strain response, b) the inverse square root density as a measure of homogenized line length, i.e., $L_\rho = 1/\sqrt{\rho}$, c) the scale parameter of the exponential distribution of line lengths, i.e., $\lambda$ for each fit of $P(L) = \lambda \exp(-\lambda L)$ for the counted line length distributions at each strain value, and d) the mean and median values of the line length distributions showing both median (darker colors) and mean (lighter colors) for the cases of equal (solid lines) and unequal (dashed lines) Peierls stress for edge and screw components.

Figure 8 also shows complementary measures of length in subplots c and d. Based on the available segment structured output, the length distributions for the present simulations were calculated, along with the homogeneous density estimate, as a function of strain. In agreement with the findings of Ref. [32], the present length distributions tended to follow an exponential distribution:

$$P(L) = \lambda \exp(-\lambda L) \tag{9}$$

where the probability density $P$ is a function of the segment length $L$ and a scale parameter $\lambda$. Examples of the exponential fit can be found in the supplemental information. The scale parameters of the fitted exponential distributions are plotted in Figure 8c, with larger numbers indicating a tighter distribution of shorter segments. Similar information is seen more directly in the comparison of both mean and median shown in Figure 8d. Based on either subplot, the dislocations of the unequal case are seen to be slightly longer than for the equal case.

The higher hardening rate, corresponding to the unequal case compared to the equal case, for the longer line length distributions shown in Figure 8c and d is contrary to intuition based on the bowing stress of a pinned dislocation being inversely proportional to its length (i.e., $\tau = \mu b / l$). However, the increase in the homogeneous line length estimate (see Figure 9b) corresponds to a drop in dislocation density, which would explain the lowered strength for the case of equal Peierls stresses. The relatively high dislocation density for the case of unequal Peierls stresses was common among the configurations simulated here. The dislocation density distribution at an axial strain of 0.9%, the largest common to all simulation runs, is shown in Figure 9 with the same numbering convention as before.

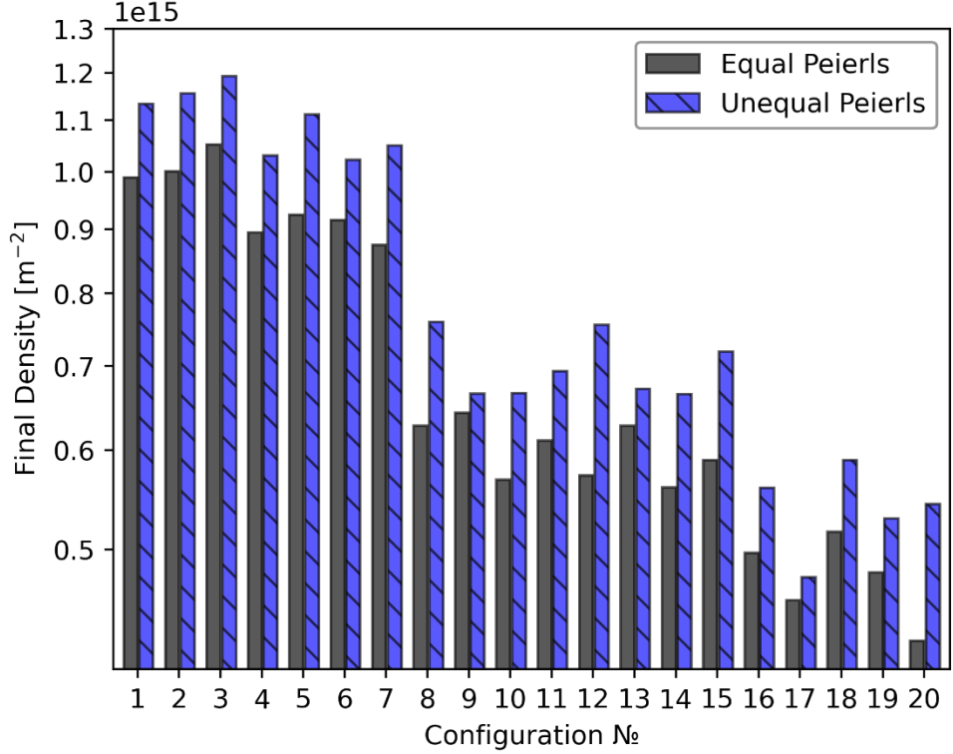

Figure 9. Distribution of dislocation densities among all simulations at an axial strain of 0.9%.

An additional theoretical framework through which to view the relation between hardening rate and line length distribution comes from a combination of dislocation density and mean free path. Ref. [4] outlines an argument that will be summarized here, beginning with a square root dependency of strength on dislocation density:

$$\tau_d = \alpha \mu b \sqrt{\rho} \tag{10}$$

where $\tau_d$ is the strength contribution due to forest hardening, $\alpha$ is a material parameter, $\mu$ is the shear modulus, $b$ is the Burgers vector magnitude, and $\rho$ is the dislocation density. The evolution of dislocation density with shear strain is then related to the motion of dislocations as

$$\frac{\mathrm{d}\rho}{\mathrm{d}\gamma} = \frac{\mathrm{d}l}{b\mathrm{d}A} = \frac{1}{b\Lambda} \tag{11}$$

where $l$ is the length of a dislocation that sweeps out a plastic area $A$, and $\Lambda$ is its mean free path. Then, differentiating Eqn. 10,

$$\tau_d \frac{\mathrm{d}\tau_d}{\mathrm{d}\gamma} = \frac{(\alpha\mu b)^2}{2} \frac{\mathrm{d}\rho}{\mathrm{d}\gamma} \qquad (12)$$

Substituting back in Eqn. 11 and defining the hardening rate as $\Theta \equiv \mathrm{d}\tau_d/\mathrm{d}\gamma$ gives

$$\tau_d \Theta = \frac{(\alpha\mu)^2}{2} \frac{b}{\Lambda} \qquad (13)$$

Rearranging, one obtains

$$\frac{\Theta}{\mu} = \frac{\alpha}{2} \frac{L_\rho}{\Lambda} \qquad (14)$$

If both $L_\rho$ and $\Lambda$ are treated as functions of character angle, the contributions to overall hardening rate can be separated by dislocation character, and the hardening rate is proportional to their ratio:

$$\Theta(\theta) \propto \frac{L_\rho(\theta)}{\Lambda(\theta)} \qquad (15)$$

The distributions of the above dislocation character dependent length ratios for each of the Peierls stress cases for the example configuration is shown in Figure 10. The distribution clearly changes between the two Peierls stress cases. For the case with a higher Peierls stress for screw-type segments, the length ratios, representing abundance without much plastic activity, appear to be much higher for low character angles. These length ratio values for screw segments are high both relative to edge segements and to most of the values seen for the case of equal Peierls stress.

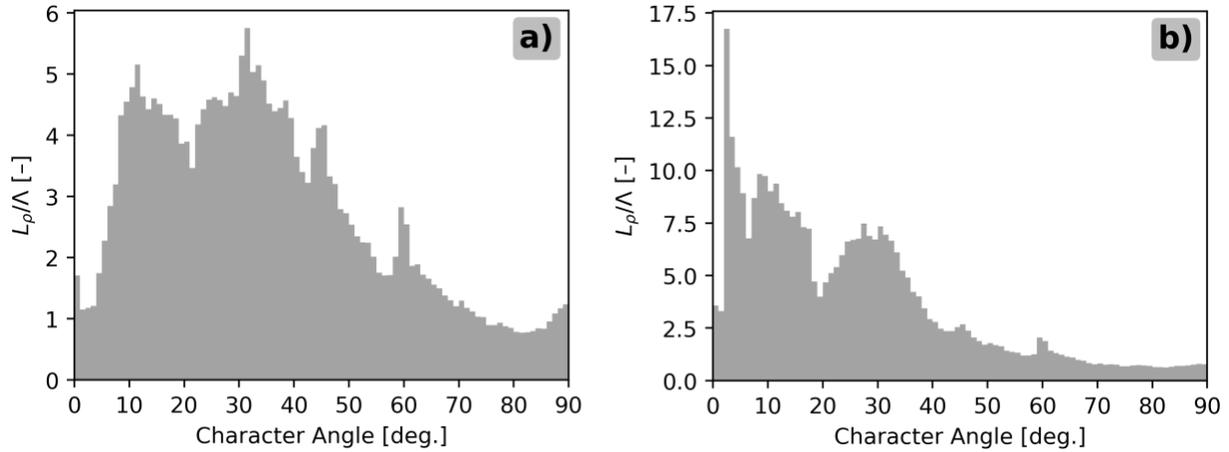

Figure 10. Using the mean homogenized length over the same strain range used to determine the hardening rate, the ratio of length over mean free path is shown here as a function of character angle. This is for the case of configuration 6 (identified in Figure 2) for a) the case of equal Peierls stress components and b) high screw Peierls stress component.

If the distributions of length over mean free path are summed over all character angles and compared to the overall hardening rates for these simulations, they slightly underpredict the increase in hardening rate between the high Peierls stress case relative to the equal Peierls stress case, suggesting a ratio of 1.35 instead of the observed ratio of 1.67. Table 3 lists these ratios for both the example case (configuration 6 in Figure 2) and the mean of all simulations. The underprediction of the observed hardening rate based on the ratio of length to mean free path is worse for the average simulation than for the example configuration. However, in both the average and specific cases, the ratio does match better with the stress level at the beginning of the strain range used for hardening rate comparisons. The strain evolution of the ratios shown in Figure 10 shows no clear trend, as discussed in the supplemental information.

Table 3. Hardening rate and length ratios for the mean of all configurations and for configuration 6 (defined in Figure 2 of the main document) for both cases of Peierls stress, where "unequal" indicates a higher value for screw components. Both quantities are taken

over the axial strain range of 0.4%–0.9%. Additionally, the stress value at a strain of 0.4% is given for extra context for the hardening rate values.

| | Quantity | Equal Value | Unequal Value | Unequal over equal ratio |
|---|---|---|---|---|
| Configuration 6 | $\Theta$ | 170. | 282 | 1.67 |
| | $L_\rho/\Lambda$ | 253 | 342 | 1.35 |
| | $\sigma_{33}(\varepsilon_{33} = 0.4\%)$ | 277 | 376 | 1.36 |
| Mean of all Configurations | $\Theta$ | 86.2 | 187 | 2.10 |
| | $L_\rho/\Lambda$ | 179 | 236 | 1.32 |
| | $\sigma_{33}(\varepsilon_{33} = 0.4\%)$ | 270. | 347 | 1.29 |

## Strengthening Superposition

In the preceding analysis, the total strength was taken to be a function of the dislocation density only. Typically, hardening contributions considering different strengthening effects (e.g., lattice friction, forest, solute, precipitate strengthening effects) are superposed to provide an estimate of the cumulative hardening. In the context of the present work, consider the following superposition expression for the total strength $\tau$

$$\tau = (\tau_d^n + \tau_P^n)^{\frac{1}{n}} \tag{16}$$

where $\tau_d$ is the dislocation density strengthening and $\tau_P$ is the contribution from the intrinsic material resistance, which includes the Peierls stress. Starting with $n = 1/2$ and the relation for $\tau_d$ in Eqn. 10, the hardening rate becomes

$$\frac{d\tau}{d\gamma} = 2\left(\tau_d^{\frac{1}{2}} + \tau_P^{\frac{1}{2}}\right)\left(\frac{1}{2}\tau_d^{-\frac{1}{2}}\frac{d\tau_d}{d\gamma} + \frac{1}{2}\tau_P^{-\frac{1}{2}}\frac{d\tau_P}{d\gamma}\right). \tag{17}$$

Since $\tau_p$ is a property of the whole network, $\frac{d\tau_P}{d\gamma} = 0$ and

$$\frac{d\tau}{d\gamma} = \left(\tau_d^{\frac{1}{2}} + \tau_P^{\frac{1}{2}}\right)\tau_d^{-\frac{1}{2}}\frac{d\tau_d}{d\gamma}, \tag{18}$$

which can be rearranged into

$$\Theta \equiv \frac{d\tau}{d\gamma} = \left(1 + \sqrt{\frac{\tau_P}{\tau_d}}\right)\frac{d\tau_d}{d\gamma}. \tag{19}$$

The extra hardening from the increased Peierls stress considered in the present work would then be, relative to the equal case:

$$\frac{\Theta_{uneq}}{\Theta_{eq}} = \frac{\left(1 + \sqrt{\frac{\tau_P^{uneq}}{\tau_d}}\right)}{\left(1 + \sqrt{\frac{\tau_P^{eq}}{\tau_d}}\right)}. \tag{20}$$

For the friction stress contribution, $\tau_P = 9$ MPa for the equal case and for the unequal case, the contribution should be within the range of Peierls stress values, i.e. between 9 and 300 MPa. For an estimate of the dislocation density strengthening contribution to the hardening rate, the stress at the beginning of the region used to calculate hardening rate was decomposed by Eqn. 16. Using the equal Peierls stress case and taking the stress value of 277 MPa (as listed in Table 3), Eqn. 16 estimates the dislocation density contribution as $\tau_d \approx 186$ MPa. Then, setting $\tau_P^{eq} = 9$ MPa and $\tau_P^{uneq} = 300$ MPa (i.e., the limiting values) results in an expected hardening ratio of 1.86. Thus, the observed hardening ratio of 1.67 between the Peierls stress cases for the example configuration discussed so far falls within the expected range. Importantly, this contribution is much closer to the screw Peierls stress value (300 MPa) than the edge value (9 MPa), which again suggests the outsized effect of the increased screw dislocation friction stress on the overall strength of the network.

Repeating the calculation of Eqn. 20 for all cases gives a value of 1.87 for the mean increase in hardening rate due to the high screw Peierls stress. However, the ratios are tightly clustered, ranging from 1.81 to 1.97 and with an interquartile range of 1.85 to 1.89. In the simulations, due to the stochastic nature of the dislocation network development and the possibility to have a negative hardening slope in some simulations with low dislocation density, the hardening ratio values ranged from -28 to 34. However, their interquartile range of 0.95 to 2.8 falls more in line with the predictions. Overall, this analysis indicates that the superposition principle proposed in Eqn. 16 reasonably predicts the hardening due to forest interactions and anisotropy in the Peierls stress.

*Graph Analysis Metrics*

The connected structure of the length output may also be analyzed to help clarify the increased initial hardening rate of the high screw Peierls stress simulations. There are many algorithms to deal with graph structured data, and while few have an intuitive application to dislocation networks, several have the potential to describe some general qualities of the structure of the dislocation graph. For example, the inverse of the number of edges between two nodes (in the present context, these nodes would be two contiguous segments in the network) makes up a measure of efficiency of the graph [33], which has been used to discern the structure of networks whose connectivity falls between fully regular and fully random [34]. Here, the unweighted efficiencies were calculated, which do not consider any lengths or spatial distances. Therefore, higher efficiency values indicate more densely connected segments but do not describe the length of the segments.

Both the global and local variants of efficiency are included in Figure 11, which also includes a count of the total number of segments and the number of connected components. The values of global efficiencies shown in Figure 11b are much lower (~0.09) than those seen for "small-world" networks of internet communication structures and transportation systems (~0.4); however, the values of local efficiency in the present context (~0.4) are comparable to those applications (~0.3–0.4) but far from, e.g., the structural connections between regions within the cerebral cortices of cats (global efficiency 0.69, local efficiency 0.83) [33]. Overall, the high local efficiencies in the dislocation networks indicate that neighboring segments tend to connect to each other in multiple ways – as an example, notice the high connectivity within the binary junction in Figure 3. The low global efficiency values indicate that connections between any two random segments must be made through a long series of neighbors, as no shortcuts exist in the network. While the efficiency values do not match the high values characteristic of small-world networks, the use of efficiencies, particularly those weighted by a suitable edge property, may be useful in future descriptions of network configurations.

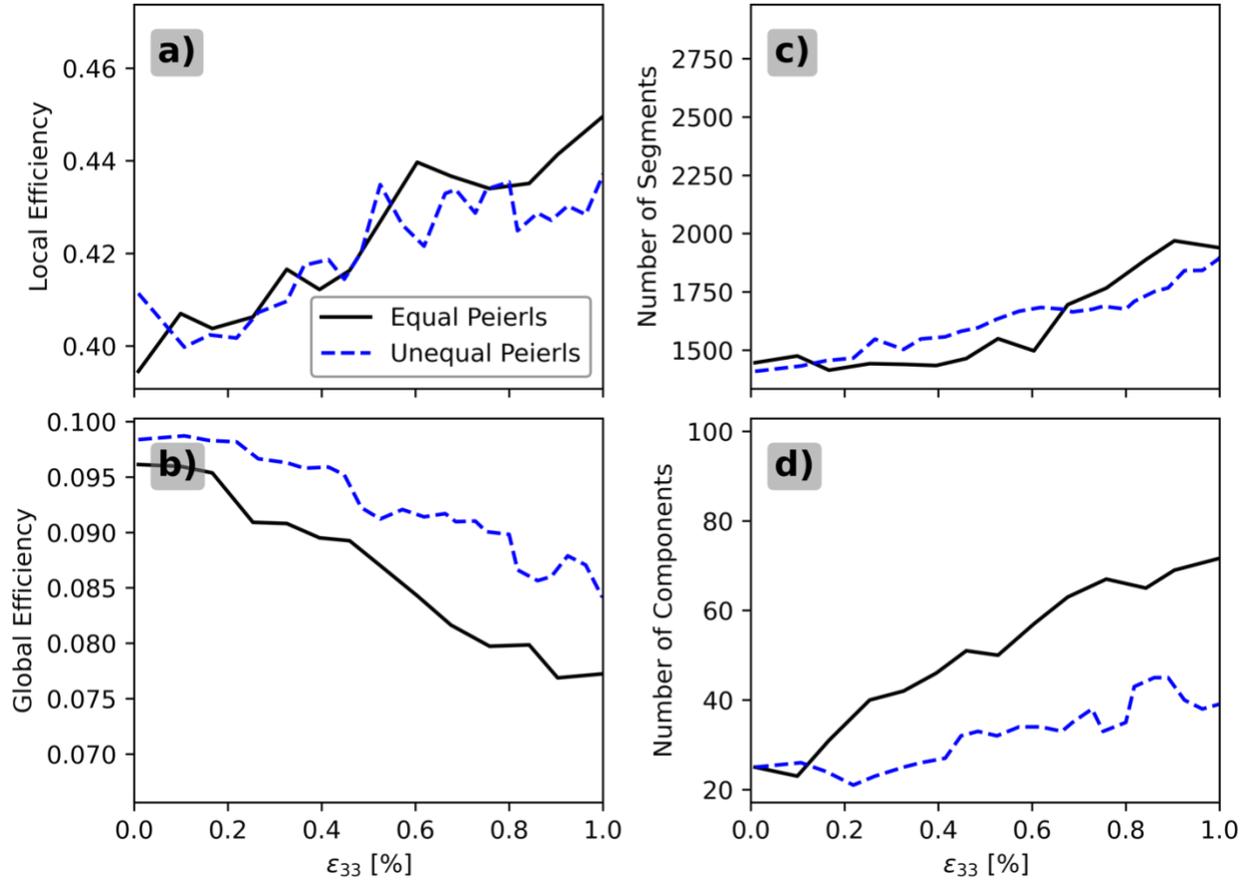

Figure 11. Graph-based metrics for configuration 6 (defined in Figure 2) showing a) local efficiency, b) global efficiency, c) number of segments, and d) number of connected components.

As illustrated in the schematic of Figure 3, a connected component in this sense would be any group of dislocations that connect to each other. The entire dislocation network may be composed of one or many connected components, but the present simulations all have dozens (see Figure 11d), with the number increasing with deformation. The rise in the number of connected components over the deformation range describes the dislocation population separating into disconnected sections. This could occur by, e.g., the unzipping of junctions, but the exact mechanisms are not contained in the network connectivity data. By any mechanism, the higher number of connected components for the unequal case indicates that the increased screw component Peierls stress served to keep the network more connected throughout deformation.

Based on Figure 11, the present network representations show a trend with increasing strain for an increase in local efficiency and a decrease in global efficiency. The increase in local efficiency may be linked with an increase in junction density, which, as in the schematic in Figure 3, would have relatively high local efficiencies. The decrease in global efficiency with strain may be explained by the rise in the number of connected components, which results from the dislocation population separating into disconnected sections. Both global efficiency and number of connected segments show a difference between the case of equal and unequal Peierls stress, with the unequal case having a lower number of connected components and a higher global efficiency. In contrast, despite showing a trend over strain, neither local efficiency nor the total count of segments distinguishes between the two cases.

Discussion

In the construction of constitutive models of slip system strength evolution for use in crystal plasticity modeling, the total dislocation density is often partitioned into separate populations [18]. For example, the total density may be separated into categories of mobile and sessile natures [35], forest and debris segments [36], or edge and screw characters [37]. The model of Arsenlis et al., for example, includes a term for the intrinsic lattice resistance to deformation in the evolution equations for edge and screw populations separately [37]. The present results, however, suggest that the properties of those populations are not separable in their effects on the overall network hardening response. Here, a change in the properties of screw dislocation segments resulted in a change in the connectivity of the entire network, with only isolated effects on the mean free paths of segments with characters very near to pure screw type. Therefore, either an additional description of the network structure itself or a more general treatment of mixed-character dislocation density evolution is necessary to translate plastic flow activity within crystal plasticity formalisms. Further, the cumulative strengthening effect in such scenarios is likely best described by a functional form that is more complex (i.e., non-linear) than a simple linear sum of strengthening effects.

The notable change in connectivity between Peierls stress cases, as seen in the divergence in the number of connected components in Figure 11d, might also have implications for the diffusion of point defects, which occurs more quickly along dislocation lines (i.e., pipe diffusion) and is influenced by the strain field of a dislocation network [38–40]. The ease of diffusion through a dislocation network may differ depending on how its connectivity changes with deformation. Note that to connect diffusivity to graph efficiency values, however, the dislocation segment graph should be weighted by real space distance. Additionally, many other graph-based metrics are available to analyze the structured output, some of which are shown in the supplemental information.

On a final note, previous studies have begun to explore graph formulations of dislocation networks, but in the work of Starkey et al., for example, the pinning points of a bowed dislocation represent nodes between which a dislocation spans as a connecting edge [41]. The present formulation is the inverse: contiguous dislocations are stored as nodes and connected amongst each other by edges. The motivation here is to store connectivity and nodal information of the segments, while the work of Starkey et al. advances a reformulation of the dynamics [41]. Overall, both the present study and the work of Starkey et al. demonstrate the utility of revisiting the analysis of dislocation networks and microstructures through the perspective of graph formulations.

Conclusions

The impact on initial hardening rate from dislocation character angle-based anisotropy in the Peierls stress of pure Ni was explored through DDD simulations of two cases, one of equal edge and screw components and one with an increased screw component. Pairs of identical initial dislocation configurations were simulated in uniaxial tension up to a strain of at least 0.9%. Analysis based on density, line length, and connectivity followed, with special attention on potential connections to hardening rate. The following conclusions can be drawn.

- Average initial hardening rate was found to be about twice as high in the cases with the anisotropic Peierls stress imposed (i.e. screw component increased by a factor of 30). The anisotropic cases also showed higher final dislocation densities relative to their initial values.

- The high screw Peierls stress resulted in the suppression of plasticity for segments only of very nearly screw type, with the rest of the character distribution remaining very similar after configuration is accounted for.

- The distributions of segment length in each dislocation network show generally longer segments for the case of unequal Peierls stress, failing to explain their higher hardening rate. Incorporating estimates of mean free path as a normalizing factor to dislocation length showed lower plastic motion for screw-type segments but slightly underpredicted overall hardening rate increases.

- A superposition principle considering forest and lattice resistance-related strengthening is proposed that provides reasonable estimates of the hardening due to these two effects.

- Structured data describing the connections between segments were analyzed with several approaches from the field of graph theory, with the most striking differences being in the number of connected components and the global efficiency, whose trends suggest that the increase in screw segment Peierls stress encouraged the preservation of more fully connected dislocation networks during deformation and may contribute to the increased hardening rate.

Acknowledgements

D.B. and L.C. acknowledge support from the U.S. Dept. of Energy, Office of Basic Energy Sciences Project FWP 06SCPE401. JDS acknowledges support from the Department of Energy National Nuclear Security Administration Stewardship Science Graduate Fellowship, provided under cooperative agreement number DE-NA0003960. Computations were performed on The Pennsylvania State University's Institute for Computational and Data Sciences' Roar supercomputer as well as on Bridges-2 at the Pittsburgh Supercomputing Center, supported by the ACCESS (previously XSEDE) program of the National Science Foundation.

Supplemental Information for:

Effect of Anisotropic Peierls Barrier on the Evolution of Discrete Dislocation Networks in Ni

John D. Shimanek[1], Darshan Bamney[2], Laurent Capolungo[2], Zi-Kui Liu[1], Allison M. Beese[1,3]

[1] Department of Materials Science and Engineering, The Pennsylvania State University, University Park, PA 16801, USA

[2] Materials Science & Technology Division, Los Alamos National Laboratory, Los Alamos, NM 87544, USA

3 Department of Mechanical Engineering, The Pennsylvania State University, University Park, PA 16801, USA

*Time Integration Procedure*

Macroscopic loads were imposed to the domains via a quasistatic loading scheme. The time integration step for dislocation motion was fixed at 0.1 picoseconds. However, during the uniaxial tension loading, another time stepping procedure was used to sequentially increase and hold the macroscopic strain of the cell, allowing the dislocation structure to relax at the smaller time integration frequency before the macroscopic procedure repeated. Each relaxation of the dislocations continued until the overall plasticity accomplished over the dynamics step was comparable to a value of the noise occurring in the system at even zero applied stress due to remeshing effects and trial topological reactions, both of which result in a small amount of plastic strain that was regarded as the noise floor for the simulation. After each relaxation, the macroscopic strain would be increased by an amount that was expected to increase the relevant stress component by 1 MPa. This loading procedure allowed for an estimation of the quasistatic response while retaining the fine resolution of time integration during plasticity to ensure the accurate motion and interaction of dislocations.

Due to the time stepping method used to simulate quasistatic strain rates, the dislocation time steps are unlinked from the macroscopic sense of time used to control the applied strain. Using a constant frequency of dislocation time steps, the area swept during any one step might happen to be taken from the tail end of relaxation, meaning the step would be nearly free of plasticity. Similarly, data from right after the application of a step in macroscopic strain might contain an overrepresentation of plasticity. Periodically sampled output was therefore not considered representative or comparable across time steps or between samples.

In light of the potential differences of dynamic quantities from one time step to another, a cumulative approach was implemented, demarcated by the relaxation conditions of each strain step. Values of both length and area were summed until an output step that took place at the end of a relaxation of an imposed strain increase, at which point the cumulative values were output and the variables reset. Thereby, each output contained comparable periods of plasticity. Variables not based on integrated dynamical quantities, like snapshots of the total and per system densities, were output at the fixed intervals and additionally output along with length and area metrics. As a result, the output occurred at two frequencies, one at fixed intervals of time integration steps and one tied to the dynamic behavior of the simulation. Also note that area and length values were separated by slip system and grouped into bins of one-degree widths, and they therefore contain no information on the correlation of particular dislocation segments either spatially within one time step or temporally across the duration of deformation.

*Effect of Periodic Boundary Conditions*

As mentioned in the main text, the interplay between the periodic boundary conditions and the imposed crystallographic orientation. Previous work has noted the importance of considering periodic boundary conditions but focused on the effect of non-cubic simulation cells [1]. The current simulations were initialized beginning with large circular loops whose image segments, when folded back into the main simulation cell, interacted with each other. The 1° of tilt away from [001] controlled the extent of interaction between image segments, whose glide plane was modified by that tilt every time it crossed a simulation boundary. As a result of the small tilt value, the image segments in the present simulations were situated in planes that were very close to each other. Since no climb was allowed, the only interaction between these dislocations was elastic, and they tended to form dipoles.

Figure S1 shows an example configuration where a large circular dislocation was initialized within the same cell bounds and crystallographic orientation as prescribed in the main simulations. Upon relaxation, two of the points where this dislocation crosses over on itself form dipoles, as shown in Figure S2. Without the prescribed tilt, these segments would have reacted to form separate loops. With a larger tilt, the dipoles had less of a driving force to form a dipole due to the competing consideration of their own line tensions relative to the elastic interaction from parallel slip planes that were further away than in the 1° case. With a tilt of 10°, for example, the dipole was more loosely bound and preferred a character angle of ~40° rather than the 60° dipoles seen with a 1° tilt.

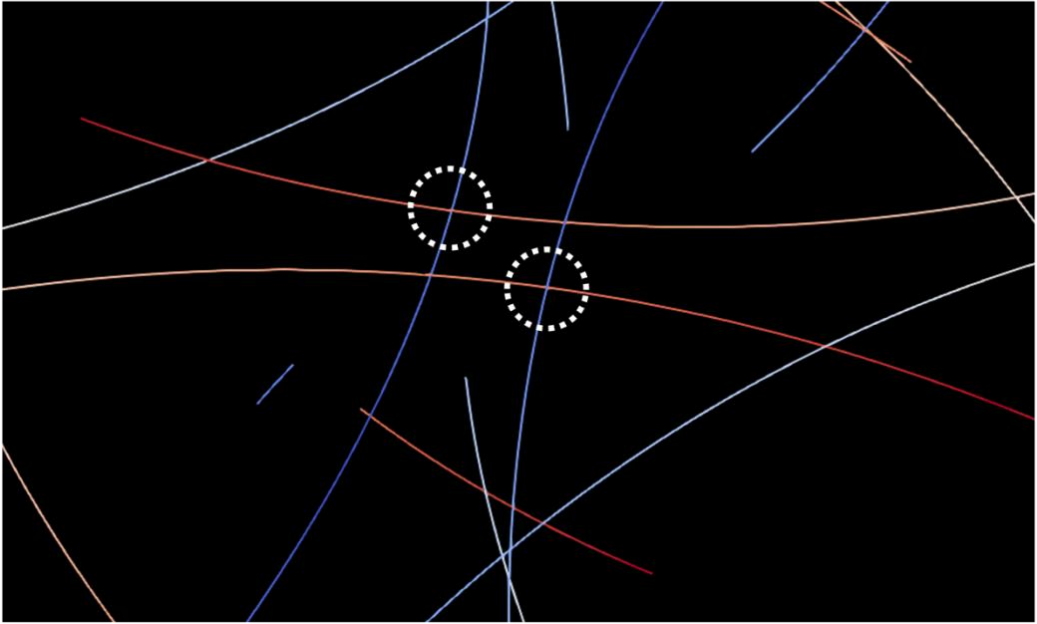

Figure S1. Large circular loop whose extension beyond the simulation cell forms image segments that closely interact with themselves. Color indicates dislocation character, with edge segments in red and screw segments in blue. In dashed circles are two instances where segments on closely spaced planes are about to interact.

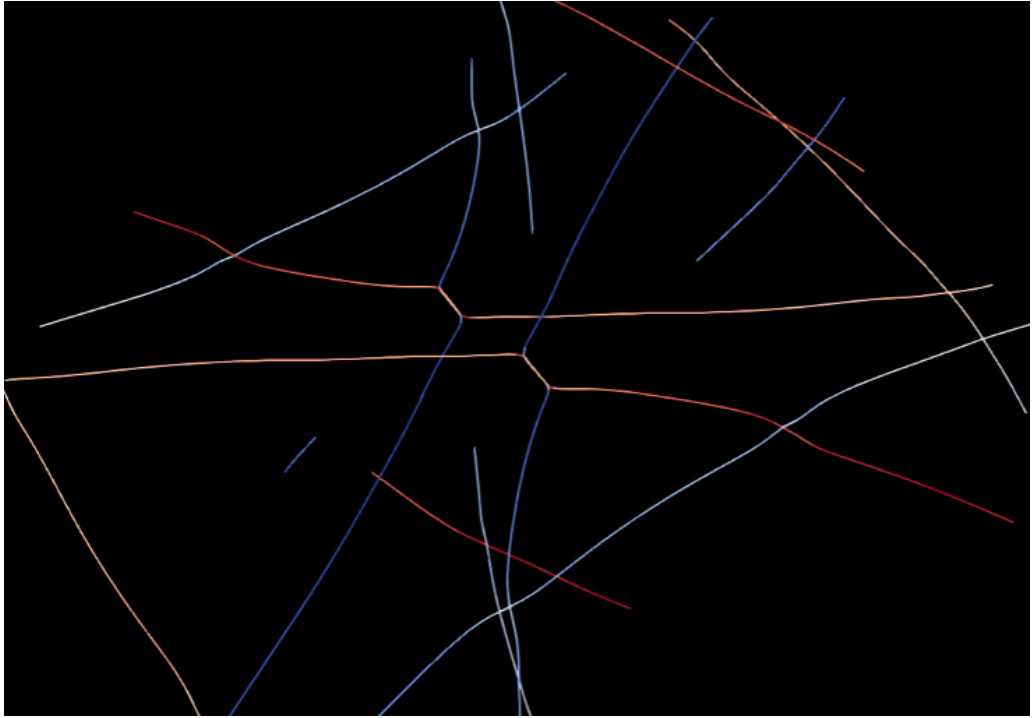

Figure S2. The same configuration as shown in Figure S1 after several simulation time steps. The segments on tightly spaced planes have formed dipoles of approximately 60˚ character.

The effect of dipole formation manifests in the tracked plasticity within the overall network due to the tendency of the dipole segments to vibrate back and forth based on their elastic interactions with each other. The swept area of the dislocation network that was tracked over the course of deformation should not account for such vibrations, and so, to mitigate the issue, the character-based metrics were recorded for each slip system. Since each system has an arbitrary but fixed sign associated with it, the plastic area per slip system does not accumulate area from back and forth motion. The per system values were later recombined to give the length and plastic area quantities used in the main document. However, the vibration of the dipole segments may still have contributed to the large plasticity seen around 60˚ in the main document, despite the strategy to first separate the metric by system.

*Graph Storage*

During each simulation, the data structure describing the dislocation network properties and connectivity was appended to a plain text file formatted in graph markup language (GML), which can be read by the NetworkX Python package [2], later used for analysis. The output was saved for analysis in a hierarchical data format (HDF5) after a per-graph compression (using Python's zlib, which enabled a compression factor of around 7) of the GML byte stream. Each graph was indexed by simulation name and, at a level below that, by simulation time. Standard outputs of stress state, strain state, dislocation density, and per system lengths and areas were scraped into a similarly structured but separate database for later analysis.

*Additional Graph Metrics*

Several additional graph metrics may be considered in their potential for describing the dislocation network evolution, among them measures of centrality, assortativity, and clustering. Here, the centrality refers to the mean fraction of nodes to which each node is connected. As shown in Figure S3a, degree centrality decreased on average, in line with trend of disconnection shown in the rise in the number of connected components in Figure S3d. Next, the assortativity of a graph quantifies the tendency of similar nodes to be connected. Here, the relevant similarity is defined in terms of the connectivity of each node [3]. Figure S3b shows that no real trend is apparent over the course of deformation and that both Peierls stress cases show similar values. Finally, the clustering of one node of a graph describes the fraction of its neighbors that connect to each other normalized by the total possible number of such connections; the exact quantity is computationally intensive but can be estimated by a sampling method [4]. Figure S3c shows a slight increase in the clustering coefficient as a function of strain but no significant difference between the Peierls stress cases. The quantity shown in Figure S3d is the mean eigenvalue of the graph

Laplacian; a higher value may be interpreted as more connectivity, but it and the spectral approach to graph interpretation are discussed more in the supplemental information.

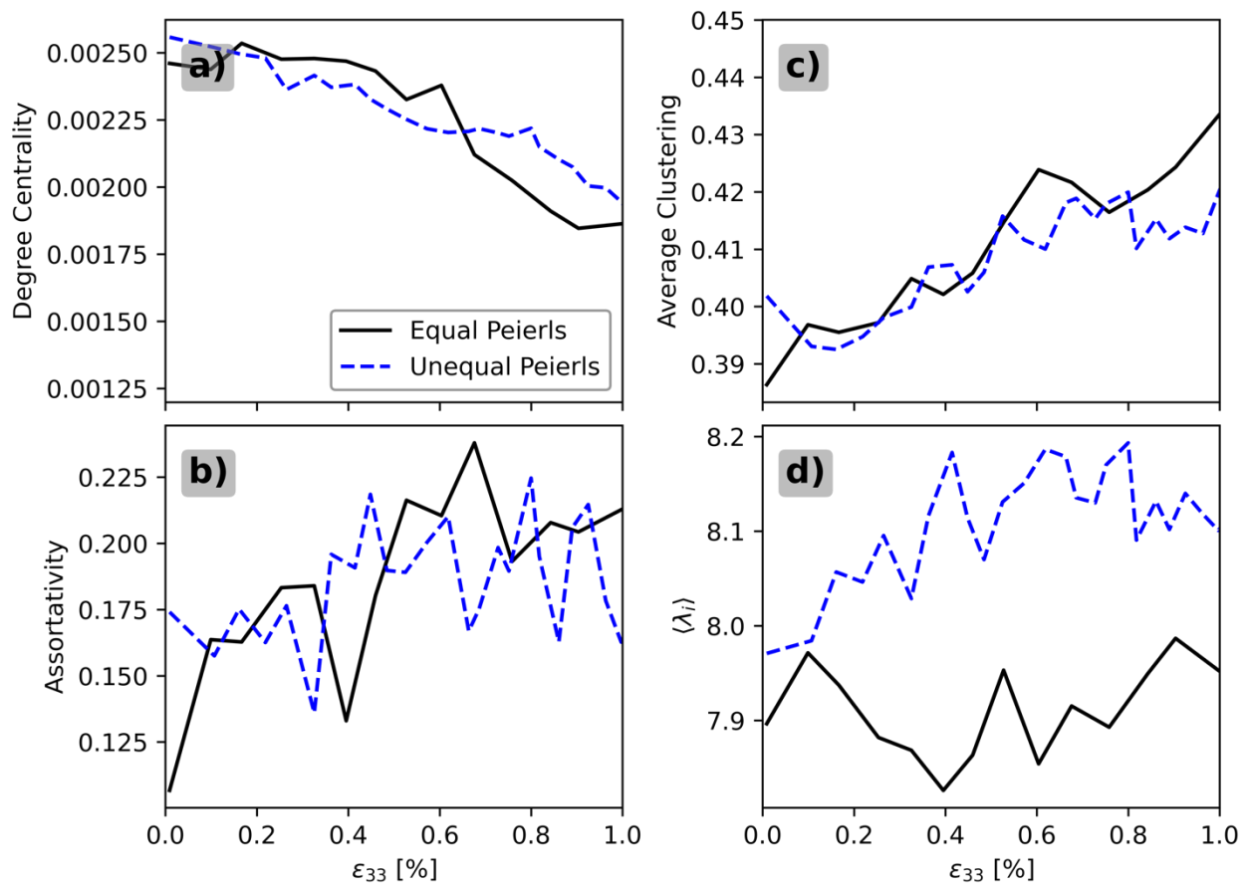

Figure S3. Additional graph-based metrics showing a) degree centrality, the mean fraction of nodes to which each node is connected, b) assortativity, the similarity of node connectedness, c) average clustering, a measure of local connectivity, and d) the mean eigenvalue of the graph Laplacian.

Due to the size of each graph, typically around one or two thousand nodes, some graph analysis methods of interest were not explored. Estimations of the graph connectivity, a measure of how the removal of individual nodes can result in disconnections within a connected subgraph [5], were computationally infeasible to run on the size of the current data representations. An analysis of cycles and graphlet motifs, as used in the analysis of other structured data [6,7], were similarly impractical. Additionally, the present analysis was limited to unweighted graphs, with no sense of physical length scale. Many of the analysis techniques could be repeated using edges weighted by some quantity of interest, perhaps related to length, curvature, or an angle of the dislocation interaction.

*Spectral Approach to Graph Analysis*

A measure of general connectivity is the algebraic connectivity, which is defined as the first non-zero eigenvalue of the graph Laplacian, a matrix describing the how each node is connected to every other node [8]. The Laplacian of a disjoint graph can be constructed by stacking those of its connected components into a block matrix [9], and, as a result, the dislocation networks analyzed here have many individual algebraic connectivity values, each relating to one connected component. However, the remaining eigenvalues in the set still encode information on the connectivity of each graph. The degeneracy of zeros, for example, is equal to the number of connected components, and the maximum eigenvalue is influenced by the number of nodes with the highest numbers of connections. Using this full spectrum, the utility of eigenvalue distributions has been leveraged in standard graph comparison methods based on the adjacency matrix [10], and as a proxy of the full Laplacian spectrum, to enable visualization over a range of strain values, its mean value was plotted in Figure S3. The mean value allows for the influence of both disconnected components and highly connected nodes to give a general estimate of connectivity for the disjoint graphs of the observed dislocation networks. The mean Laplacian eigenvalue, $\langle \lambda_i \rangle$, shows a separation between the equal and unequal cases and otherwise no obvious trend over the strain range. However, the separation could stem from the same source as shown in Figure 11d in the main text, i.e., the increase in the number of connected components for the case of equal Peierls stress. The similarity in the low eigenvalue spectrum, shown as a cumulative distribution in Figure S4, shows this not to be the case. However, the differences in the full spectrum probed at the either low or high strain, appear very similar, and no strong conclusions can be drawn.

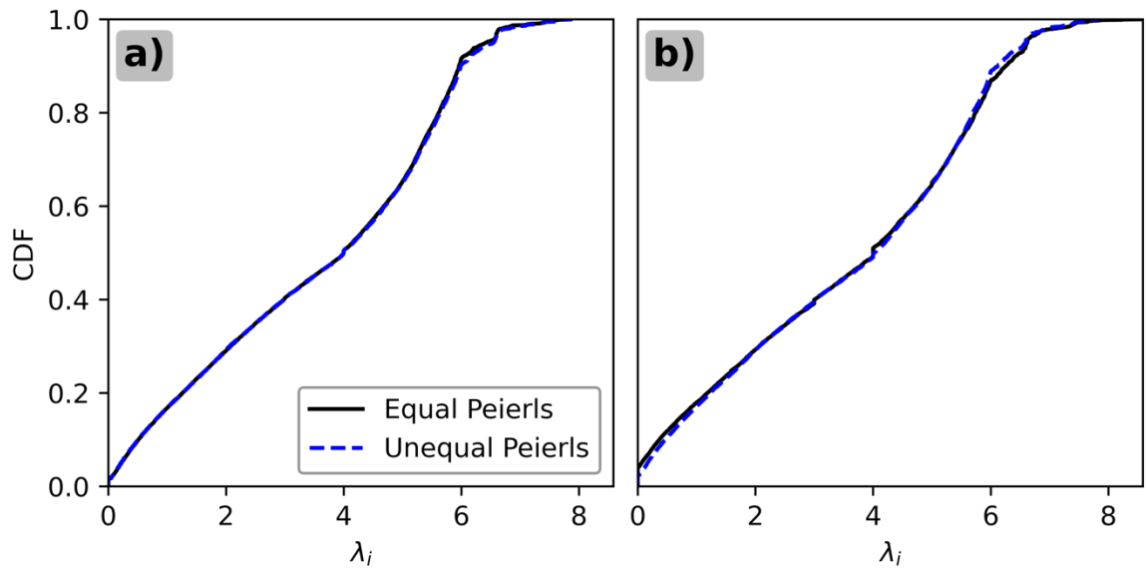

Figure S4. Cumulative distribution functions of the Laplacian eigenvalues for the example configuration at a) a strain of 0% and b) a strain of 0.9%.

Line Length Distributions

The exponential fit whose single parameter is reported as a function of strain in the main text (see Figure 8c) may also be explored at any particular value of strain probed in the simulation. Below, in Figure S5, the raw data for line lengths and an accompanying exponential fit are shown in terms of the cumulative distribution function (CDF) at several values of strain throughout deformation. In Figure S5, these values are for the case of equal Peierls stresses; the same type of analysis for the case of equal Peierls stress is shown in Figure S6.

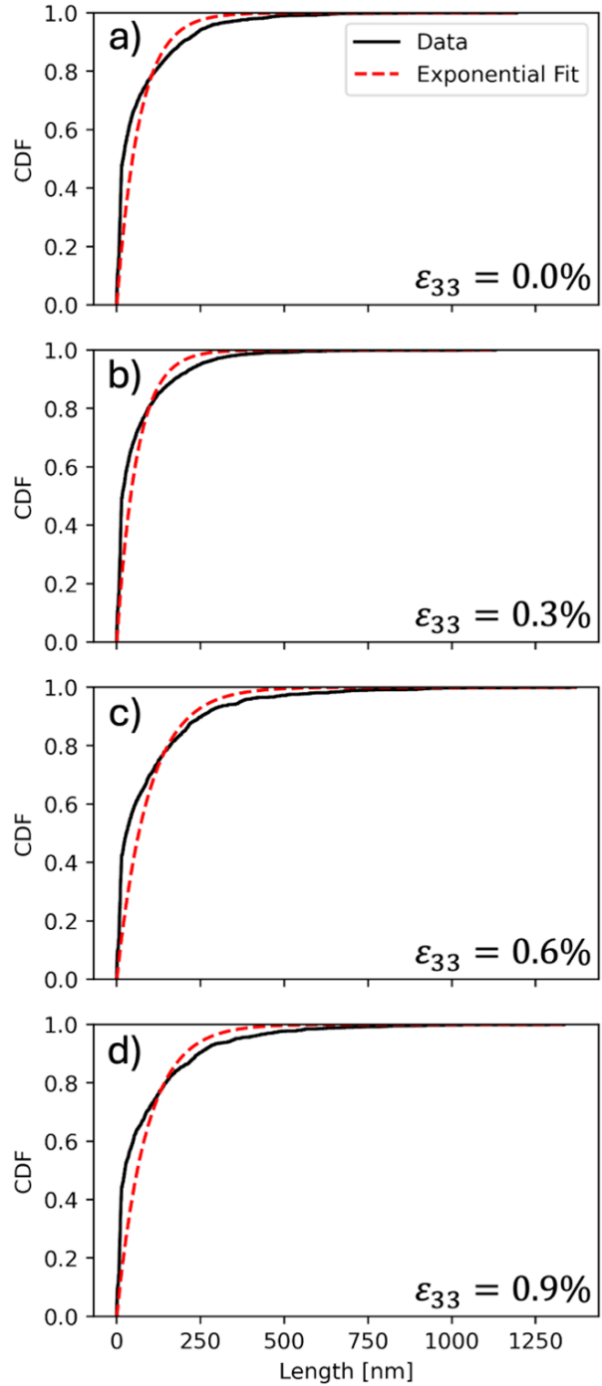

Figure S5. Line length distributions of all segments from the example run labeled as number 6 in Figure 2 of the main text using equal Peierls stresses for edge and screw components. The distributions are given as cumulative distribution functions at several strains: a) 0.0%, b) 0.3%, c) 0.6%, and d) 0.9%.

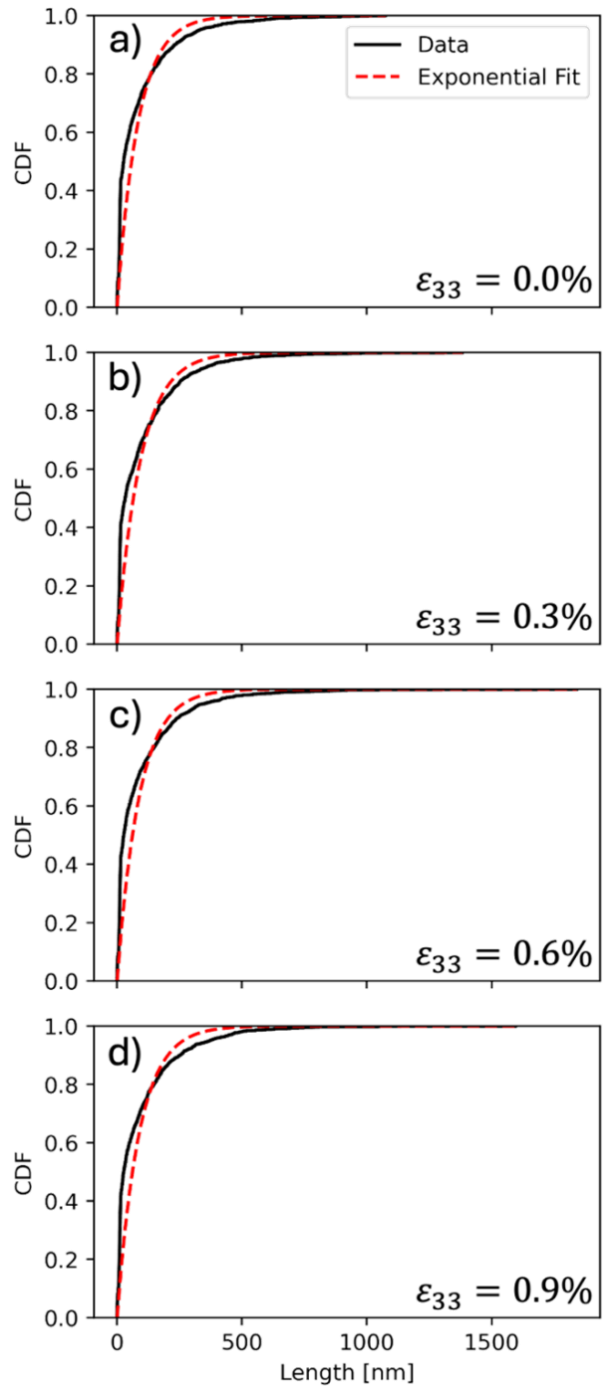

Figure S6. Similar to Figure S5 except for the case of unequal Peierls stress for edge and screw components.

From the viewpoint of Eqn 14 of the main text without dislocation character dependence, the increased hardening rate for the unequal Peierls stress case cannot be directly attributed to the larger values of segment length seen for the case of high screw Peierls stress without also considering the mean free path. Using both the homogenized length definition, $L_\rho$, and the median value of the calculated length distribution, $L$, the length ratios (i.e., length relative to mean free path, as in Eqn 14) are shown in Figure S7 along with the summed mean free path values. An overlap is seen in length ratios between equal and unequal Peierls stress cases, regardless of which definition of length is used in the numerator.

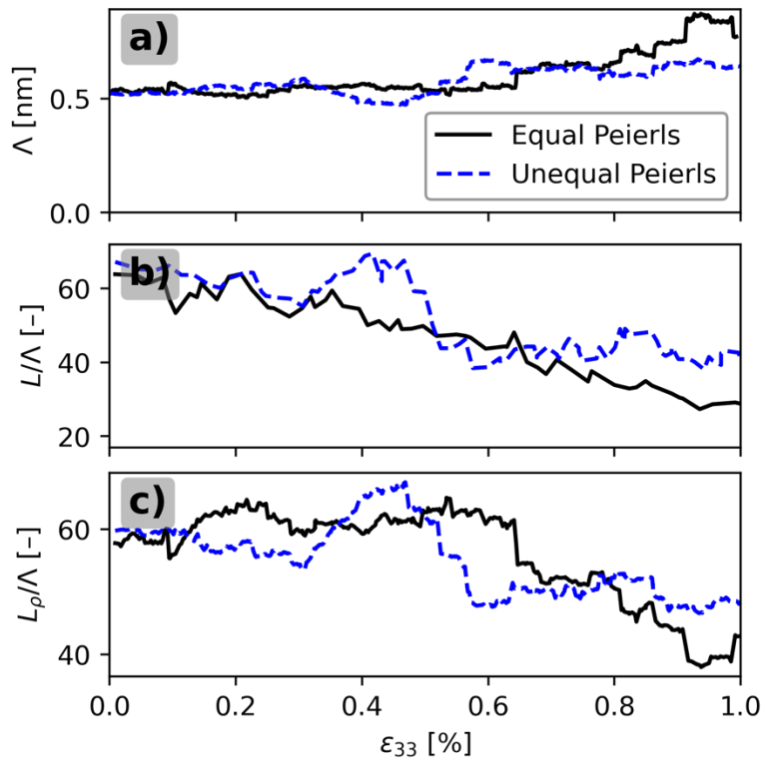

Figure S7. Evolution of several metrics describing length scales in comparable initial dislocation networks (configuration 6 in Figure 2 of the main document), showing a) the overall mean free path, b) the ratio of median length to mean free path, and c) the ratio of homogenized line length, calculated as the inverse square root dislocation density, to the mean free path.